\documentclass[preprint, nofootinbib]{revtex4}
\usepackage{amsmath, amsfonts, amssymb,graphicx, dcolumn}

\begin{document}

\title{Steady-State Two Atom Entanglement in a Pumped Cavity Enhanced by Nonlinear Mirrors}
\author{Hideomi Nihira}
\email{nihira@optics.rochester.edu}
\author {C. R. Stroud, Jr.}
\affiliation{The Institute of Optics, University of Rochester,
Rochester, NY 14627, USA}
\date{\today}

\begin{abstract}

We demonstrate the steady-state entanglement of two two-level atoms
inside a pumped cavity with photon leakage through a nonlinear
mirror and through spontaneous decay, and show that the entanglement
is enhanced by the presence of a nonlinear mirror. Our model assumes
the vacuum Rabi splitting of the dressed states of the system to be
much larger than any of the decay parameters of the system.  We also
discuss how the dressed states of the system offer us intuition as
to where the entanglement lies in the state space spanned by the
system, and allow us the optimize the system.

\end{abstract}
\maketitle

\section{Introduction}

In an earlier paper \cite{Nihira} we showed that two two-level atoms
placed in a pumped high-Q cavity with normal weakly lossy mirrors
will not have sustained entanglement in the case of closed two-level
systems, but may have such entanglement in the case of open
two-level systems. Our treatment included cavity leakage and
spontaneous emission.  It assumed that the Q-factor of the cavity
was large enough so that the vacuum Rabi splitting was larger than
any of the decay rates of the system.  This allowed us to express
the system in terms of the atom-field dressed state picture.  The
coherence of the system was built into the dressed states allowing
us to treat the effects of cavity losses and spontaneous emission in
terms of simple incoherent transitions between these dressed states.
 The advantage of expressing the system in the dressed state basis is
that we can immediately see within which manifold the entanglement
between the atoms lies, and tailor our system to maximize the
population in that manifold.

Due to experimental advances in atomic traps and cavity QED
\cite{Wienman,Wineland,Haroche, Walther, Kimble, Walther2} it is
within our technological limits to trap and manipulate individual atoms inside
a microcavity to study their entanglement behavior.  As a result, more attention is
directed towards the entanglement of atoms and fields within these cavities \cite{Guo, Li, Orozco,
Knight,Pelino, Vogel}.  More specifically, to determine the conditions in which atoms
can get entangled in these system, and to characterize the states of these entangled atoms.
One of the interesting things to note in open systems, such as the one we describe in this paper, is
that it allows the possibility of steady-state entanglement \cite{Orozco, Pelino} without the
assistance of post-selection.

The intuition gained from the dressed-state formalism suggested that
the addition of a nonlinear mirror to the cavity might allow the
production of steady-state entanglement even in the case of closed
two-level systems.  In this paper, we will demonstrate the
possibility of steady-state entanglement between two atoms by
employing a nonlinear mirror to construct the cavity.  The
nonlinearity we are interested in is the reverse-saturated
absorption (RSA) property of the mirror \cite{Malcuit, Agarwal,
Tang} which offers a larger cavity photon loss at greater
intra-cavity field intensities.  The RSA mirror effectively changes
the photon number distribution in the cavity which will
preferentially sustain the low photon number states which in turn
means a larger population in the lower manifolds of the dressed
states of the system we consider.  This concentration of population
in the lower manifolds yields the entanglement between the two atoms
in the cavity.

Here we will investigate the steady-state entanglement of two two-level
atoms inside a pumped high-Q cavity with a nonlinear mirror.  We express the state of the two atoms
as a mixed state in the dressed state basis in which the weighting factors are determined
by constructing rate equations governing the steady-state population in each of the
dressed states.  This is justified by assuming that the vacuum Rabi splitting is much larger
than any decay parameter of the system \cite{Nihira}.  We will show that only the $n=1$ manifold of the dressed states contribute
to the entanglement of the system, and how the employment of a nonlinear mirror can help
generate entanglement between the two two-level atoms inside the cavity.

\section{Model System: Two Two-level Atoms Inside a Cavity}



\begin{figure}[h]
\centering
\includegraphics[width=3in]{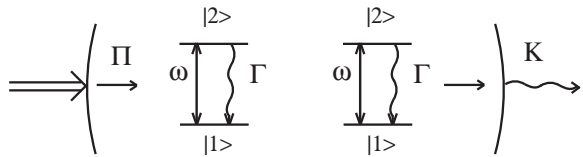}
\caption{\label{system} Two two-level atoms inside a cavity}
\end{figure}

The system we are considering consists of two atoms, each with
energy structure shown in Fig.(\ref{system}), in a cavity which is
externally pumped on resonance with the atomic transition and the cavity. The
system can lose energy through cavity leakage or through spontaneous
emission of the atoms.  However, unlike ref.\cite{Nihira}, we will take the output mirror to be
a nonlinear mirror such that the power transmission coefficient, $K$, is a function of the number of photons in the cavity.

We assume the cavity has a high Q-factor such that the vacuum
Rabi splitting is much larger than the spontaneous decay rate and the
cavity leakage rate.  The large Rabi splitting justifies the use of the rate
equations with respect to the dressed states for our system.

Since we wish to characterize this system using rate equations, we
will need the dressed states of the closed system, and calculate the transition rates between these dressed states.
From this we will construct the rate equation and determine the mixed state density matrix of the two atoms in the cavity.

The Hamiltonian of the closed system in the interaction picture is

\begin{equation}
\hat{H}_{I}=g_{1}\hat{\sigma}_{12}^{(1)}\hat{a}^{\dagger}+g_{2}\hat{\sigma}_{12}^{(2)}\hat{a}^{\dagger}+H.C.
\end{equation}

\noindent where $\hat{\sigma}_{ij}^{(n)}$ is the atomic transition
operator for the $n^{th}$ atom and $\hat{a}^{\dagger}$ is the field
creation operator.  For the sake of simplicity we will assume the coupling constants, $g_{i}$, to be the same for each atom.  In the case in which there is only one excitation in the system there are three essential states,

\begin{equation}
|11;1\rangle, |12;0\rangle, |21;0 \rangle, \\
\end{equation}

\noindent and three dressed states,

\begin{equation}
|\chi_{o}\rangle=\frac{1}{\sqrt{2}}\big(|12;0\rangle-|21;0\rangle\big),
\end{equation}

\begin{equation}
|\chi_{+}\rangle=\frac{1}{\sqrt{2}}|11;1\rangle + \frac{1}{2} \big(|12;0\rangle+|21;0\rangle\big),
\end{equation}

\begin{equation}
|\chi_{-}\rangle=\frac{1}{\sqrt{2}}|11;1\rangle - \frac{1}{2} \big(|12;0\rangle+|21;0\rangle\big).
\end{equation}

\noindent Here $|ab;c\rangle=|a\rangle_{1}\otimes|b\rangle_{2}\otimes|c\rangle_{f}$ indicates the first atom is in state $|a\rangle$, the second atom in state $|b\rangle$, and the field in state $|c\rangle$.

The $n\geq2$ excitation of the system will have a different set of
dressed states since there are four essential states in this case.
The four essential states for $n\geq2$ are,

\begin{equation}
|11;n\rangle, |12;n-1\rangle, |21;n-1 \rangle, |22;n-2\rangle.\\
\end{equation}

\noindent The dressed states are then given by,

\begin{equation}
|\phi_{o}^{n}\rangle=\frac{1}{\sqrt{2}}\big(|12;n-1\rangle-|21;n-1\rangle\big),
\end{equation}

\begin{equation}
|\phi_{o'}^{n}\rangle=\frac{1}{\sqrt{2}}\big(|11;n\rangle-|22;n-2\rangle\big),
\end{equation}

\begin{equation}
|\phi_{+}^{n}\rangle=\frac{1}{2}\big(|11;n\rangle +
|12;n-1\rangle+|21;n-1\rangle+|22;n-2\rangle\big),
\end{equation}

\begin{equation}
|\phi_{-}^{n}\rangle=\frac{1}{2}\big(|11;n\rangle -
|12;n-1\rangle-|21;n-1\rangle+|22;n-2\rangle\big).
\end{equation}

Using the prescription described in ref.\cite{Nihira}, we can obtain the rate equations governing the population in the $n=0$, $n=1$, and $n=2$ dressed state as,

\begin{equation}
\begin{array}{l}
\frac{d{P}_{g}}{dt}=\Gamma P_{s1}+\frac{K(1)}{2}P_{s1}-\Pi P_{g},\\
\frac{d{P}_{s1}}{dt}=-\Gamma P_{s1}-\frac{K(1)}{2}P_{s1}+\Pi P_{g}+\frac{3}{2}\Gamma P_{s2}+\Gamma P_{o',2}+\frac{1}{4}(K(2)+2K(1))P_{s2}+\frac{1}{2}K(2)P_{o',2}-\Pi P_{s1},\\
\frac{d{P}_{s2}}{dt}=-\frac{3}{2}\Gamma P_{s2}-\frac{1}{4}(K(2)+2K(1))P_{s2}+\frac{3}{4}\Pi P_{s1},\\
\frac{d{P}_{o',2}}{dt}=-\Gamma P_{o',2}-\frac{1}{2}K(2)P_{o',2}+\frac{1}{4}\Pi P_{s1},\\
\frac{d{P}_{o}}{dt}=0,\\
\frac{d{P}_{o,2}}{dt}=0,\\
\end{array}
\end{equation}
\noindent with

\begin{equation}
\begin{array}{l}
P_{s1}=P_{+}+P_{-},\\
P_{s2}=P_{+,2}+P_{-,2},
\end{array}
\end{equation}

\noindent where $P_{g}$ is the population of the ground state of the
system, $P_{\pm}$ is the population in the $|\chi_{\pm}\rangle$
dressed states, $P_{0}$ is the population in the $|\chi_{o}\rangle$
dressed state, $P_{\pm,n}$ is the population in
$|\phi_{\pm}^{n}\rangle$, $P_{o,n}$ is the population in the
$|\phi_{o}^{n}\rangle$, and $P_{o',n}$ is the population in
$|\phi_{o'}^{n}\rangle$.  The Einstein A coefficient of the
$2\rightarrow 1$ transition of the single atom in free space is
given by $\Gamma$, the single photon pumping rate inside the cavity
is given by $\Pi$, and the power transmission coefficient of the
cavity output mirror as a function of the number of photons in the
cavity is given by $K(n_{p})$.  Here we assume that there is no
population initially in the $|\chi_{o}\rangle$ and
$|\phi_{o}^{n}\rangle$ dressed states, and because these states do
not couple to any other states, they will not accumulate any
population at later times.

As mentioned in ref.\cite{Nihira}, truncating the rate equation to
$n=2$ will over-estimate the entanglement content between the two
atoms since there will be some population beyond the $n=2$ manifold
that does not directly decay down to the $n=1$ manifold.  To correct
for this, we will go one step further to obtain $P_{s3}$ and
$P_{o',3}$ in order to determine what fraction of the $n\geq2$
population lies in the $n=2$ manifold.  To simplify the equations,
we will assume $K(n_{p})=K(2)$ for $n_{p}\geq2$, and express $K(2)$
as $K(2)=\eta K(1)=\eta K$.  Here $\eta$ is the measure of
nonlinearity in the mirror since it tells us how much more (or less)
the cavity transmits depending on the intra-cavity field intensity.
The equations for $P_{s3}$ and $P_{o',3}$ are then given by,

\begin{equation}
\begin{array}{l}
\frac{d{P}_{s3}}{dt}=-\frac{3}{2}\Gamma P_{s3}-\frac{1}{4}K(3\eta+1)P_{s3}+\Pi P_{s2},\\
\frac{d{P}_{o',3}}{dt}=-\Gamma P_{o',3}-\frac{1}{2}K(\eta+1)P_{o',3}+\Pi P_{o',2}.\\
\end{array}
\end{equation}

\noindent Solving the above equations in the steady-state we get,

\begin{equation}
\begin{array}{l}
P_{s3}=\frac{\Pi}{\frac{3}{2}\Gamma+\frac{1}{4}K(3\eta+1)}P_{s2}\\
P_{o',3}=\frac{\Pi}{\Gamma+\frac{1}{2}K(\eta+1)}P_{o',2}.\\
\end{array}
\end{equation}

\noindent To determine the fraction of population in the $n=2$ manifold within the $n\geq 2$ manifolds we have to solve the equations,

\begin{equation}
\begin{array}{l}
P_{s2}+P_{s3}=1\\
P_{o',2}+P_{o',3}=1.\\
\end{array}
\end{equation}

\noindent Substituting the expressions for $P_{s3}$ and $P_{o',3}$
in the above equations we get,

\begin{equation}
\begin{array}{l}
P_{s2}=\alpha=\frac{6\Gamma+K(3\eta+1)}{6\Gamma+4\Pi+K(3\eta+1)}\\
P_{o',2}=\beta=\frac{2\Gamma+K(\eta+1)}{2(\Gamma+\Pi)+K(\eta+1)}.\\
\end{array}
\end{equation}

\noindent This suggests that the rate equations for the dressed states should be modified to,



\begin{equation}
\begin{array}{l}
\frac{d{P}_{g}}{dt}=\Gamma P_{s1}+\frac{K}{2}P_{s1}-\Pi P_{g},\\
\frac{d{P}_{s1}}{dt}=-\Gamma P_{s1}-\frac{K}{2}P_{s1}+\Pi P_{g}+\frac{3}{2}\alpha\Gamma P_{s2}+\beta\Gamma P_{o',2}+\frac{1}{4}\alpha (\eta+2)KP_{s2}+\frac{1}{2}\beta \eta KP_{o',2}-\Pi P_{s1},\\
\frac{d{P}_{s2}}{dt}=-\frac{3}{2}\alpha\Gamma P_{s2}-\frac{1}{4}\alpha (\eta+2)KP_{s2}+\frac{3}{4}\Pi P_{s1},\\
\frac{d{P}_{o',2}}{dt}=-\beta\Gamma P_{o',2}-\frac{1}{2}\beta \eta KP_{o',2}+\frac{1}{4}\Pi P_{s1},\\
\end{array}
\end{equation}

\noindent where we have replaced $P_{s2}$ and $P_{o',2}$ by $\alpha P_{s2}$ and $\beta P_{o',2}$ to reflect the true population decay of the $n=2$ manifold.



\noindent The steady-state solution to these rate equations is,

\begin{equation}
\begin{array}{l}
P_{g}=\mathcal{N}\beta\alpha (2\Gamma+K)(2\Gamma+\eta K)(6\Gamma+(2+\eta)K),\\
P_{s1}=\mathcal{N}2\Pi\beta\alpha (2\Gamma+\eta K)(6\Gamma+(2+\eta)K),\\
P_{s2}=\mathcal{N}6\Pi^{2}\beta (2\Gamma+\eta K),\\
P_{o',2}=\mathcal{N}\Pi^{2}\alpha(6\Gamma+(2+\eta)K),\\
\end{array}
\end{equation}

\noindent where we have defined the normalization constant,

\begin{equation}
\begin{array}{l}
\mathcal{N}=6\beta\eta\Pi^{2}K+12\beta\Gamma\Pi^{2}+6\Pi^{2}\Gamma\alpha+\Pi^{2}\alpha\eta K+2\Pi^{2}\alpha K+24\alpha\beta\Gamma^{3}+16\alpha\beta\eta\Gamma^{2} K+\\
20\alpha\beta\Gamma^{2} K+2\alpha\beta\Gamma\eta^{2}K^{2}+12\alpha\beta\Gamma\eta K^{2}+4\alpha\beta\Gamma K^{2}+\alpha\beta\eta^{2} K^{3}+2\alpha\beta\eta K^{3}+\\
24\alpha\beta\Pi\Gamma^{2}+16\alpha\beta\eta\Gamma\Pi K+8\alpha\beta\Gamma\Pi K+2\alpha\beta\Pi\eta^{2} K^{2}+4\alpha\beta\eta\Pi K^{2}.\\
\end{array}
\end{equation}

We now want to investigate the entanglement between the two atoms in
the cavity.  To do this we trace out the field component of the
density matrix and obtain the reduced density matrix of just the two
atoms.  The reduced density matrix of the two atoms is given by,

\begin{equation}
\hat{\rho}_{atoms}=P_{g}\hat{\rho}_{g}+P_{s1}\hat{\rho}_{s1}+P_{s2}\hat{\rho}_{s2}+P_{o',2}\hat{\rho}_{o',2}
\end{equation}

\noindent where

\begin{equation}
\hat{\rho}_{g}= \begin{pmatrix}
1 & 0 & 0 & 0 \\
0 & 0 & 0 & 0 \\
0 & 0 & 0 & 0 \\
0 & 0 & 0 & 0 \\
\end{pmatrix} ,
\hat{\rho}_{s1}= \begin{pmatrix}
\frac{1}{2} & 0 & 0 & 0 \\
0 & \frac{1}{4} & \frac{1}{4} & 0 \\
0 & \frac{1}{4} & \frac{1}{4} & 0 \\
0 & 0 & 0 & 0 \\
\end{pmatrix},
\hat{\rho}_{s2}= \begin{pmatrix}
\frac{1}{4} & 0 & 0 & 0 \\
0 & \frac{1}{4} & \frac{1}{4} & 0 \\
0 & \frac{1}{4} & \frac{1}{4} & 0 \\
0 & 0 & 0 & \frac{1}{4} \\
\end{pmatrix},
\hat{\rho}_{o',2}= \begin{pmatrix}
\frac{1}{2} & 0 & 0 & 0 \\
0 & 0 & 0 & 0 \\
0 & 0 & 0 & 0 \\
0 & 0 & 0 & \frac{1}{2} \\
\end{pmatrix}.
\end{equation}

In order to calculate the entanglement content we employ Wootters'
concurrence \cite{wooters} which is defined as,

\begin{equation}
C=\max(\sqrt{\lambda_{1}}-\sqrt{\lambda_{2}}-\sqrt{\lambda_{3}}-\sqrt{\lambda_{4}},0)
\end{equation}

\noindent where $\lambda_{i}$ are the eigenvalues, in descending
order of value, of the matrix $\rho\tilde{\rho}$
($\tilde{\rho}=(\sigma_{y}\otimes\sigma_{y})\rho^{*}(\sigma_{y}\otimes\sigma_{y})$).

The concurrence of $\hat{\rho}_{atoms}$ is given by,

\begin{equation}
C_{atoms}=\max\big(\frac{1}{2}(P_{s1}+P_{s2})-\frac{1}{2}\big[(P_{s2}+2P_{o',2})(2P_{o',2}+P_{s2}+4P_{g}+2P_{s1})\big]^{\frac{1}{2}},0\big).
\end{equation}

We have shown in ref.\cite{Nihira} that in the case of a linear
mirror ($\eta=1$) there is no combination of parameters which yields
a nonzero concurrence.  Would an $\eta\neq1$ yield a nonzero
concurrence?  To answer this question, first we note that the three
dressed $|\phi_{o'}^n\rangle$, $|\phi_{+}^n\rangle$, and,
$|\phi_{-}^n\rangle$ have no entanglement (i.e $C=0$) between the
atoms.  The only manifold in the dressed state picture which offers
a nonzero entanglement between the atoms is the $n=1$ manifold
dressed states $|\chi_{\pm}\rangle$ with $C=\frac{1}{2}$. Therefore,
it stands to reason that we would want to put as much population in
the $n=1$ manifold as possible.  In order to put more population in
the $n=1$ manifold we would require $\eta>1$.  This means that the
cavity experiences a larger loss of photons for higher intra-cavity
intensities. One possible way to do this would be to incorporate a
reverse saturable absorber in the cavity \cite{Malcuit, Agarwal,
Tang}.

\begin{figure}[h]
\centering
\includegraphics[width=5in]{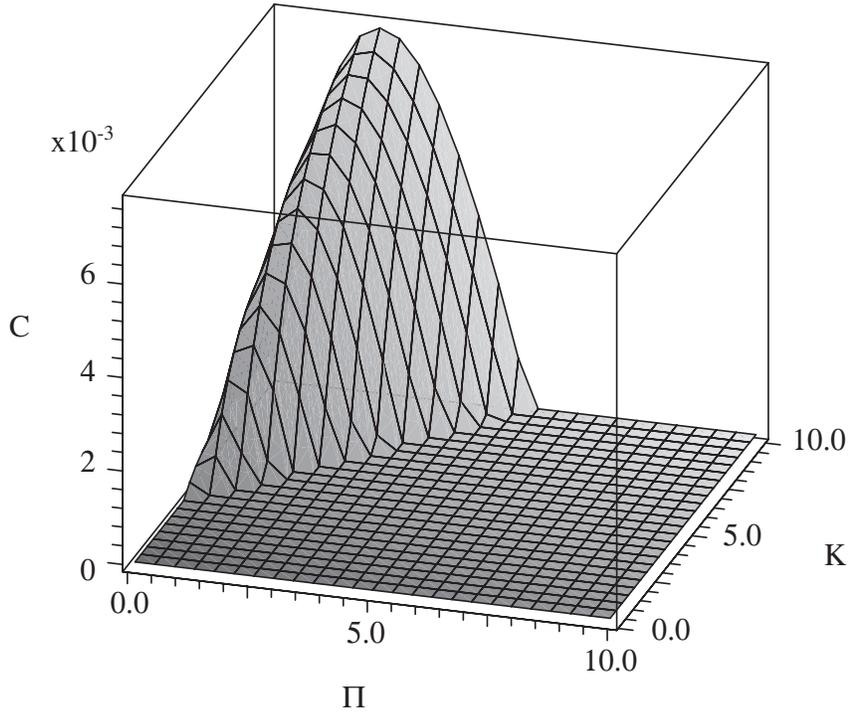}
\caption{\label{Cploteta10} Plot of C against $\Pi$ and $K$ (in
units of $\Gamma$), $\eta$=10.}
\end{figure}

\begin{figure}[h]
\centering
\includegraphics[width=4.2in]{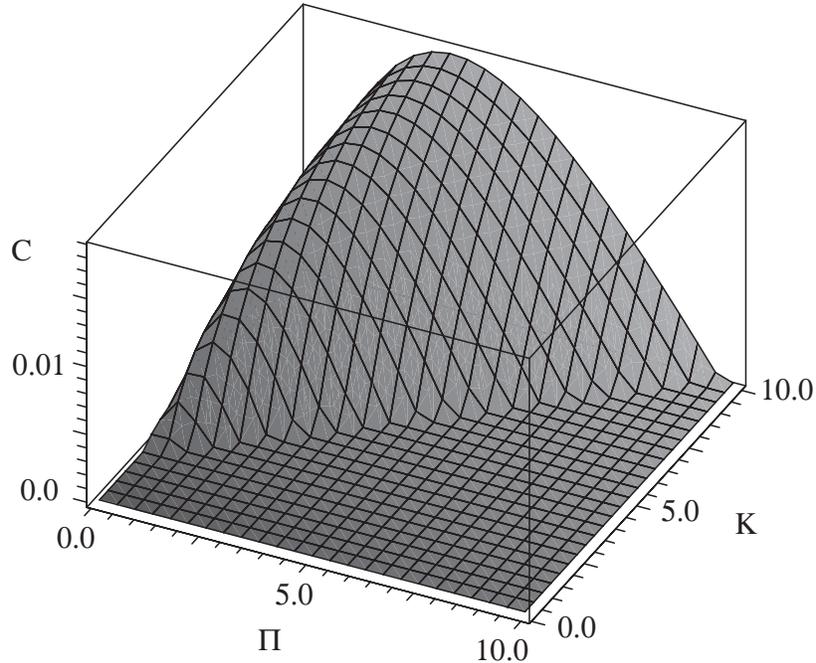}
\caption{\label{Cploteta12} Plot of C against $\Pi$ and $K$ (in
units of $\Gamma$),$\eta$=12.}
\end{figure}

We can see in Fig.(\ref{Cploteta10}-\ref{Cploteta12}) how the
entanglement between the two atoms is affected by the pump rate and
the photon leakage rate (both expressed in units of $\Gamma$) with
different values of $\eta$.  It is clear that as we increase the
value of $\eta$, the maximum value of the plot rises.  We start
seeing a nonzero value of concurrence at $\eta\approx7.746$.
However, it seems that one would need $\eta>10$ to be able to see
entanglement between the atoms for any realistic system.  This would
mean that we require a nonlinearity such that a two-photon state of
the cavity will decay at a rate which is ten times faster than that
of a single photon state of the cavity.


\section{Conclusion}

We have derived the steady-state reduced density matrix for two spontaneously decaying two-level atoms inside a high-Q cavity which is pumped and experiences photon leakage through a RSA mirror.  In our model we assumed that the vacuum Rabi splitting is much larger than any decay parameter in the system which allows us to express the density matrix of the system as a mixture of the dressed states of the system with a weighting factor that is determined by the rate equations of these dressed states.  We show that the atoms in the system can get entangled in the steady-state by choosing $\eta$, the nonlinearity parameter, to be greater than one.  Therefore, employing reverse saturable mediums can, in principle, entangle two two-level atoms inside a high-Q cavity in our model.

\end{document}